%% file: main.tex
\documentstyle[psfig]{l-aa}
\input refer
\input units

\begin{document}

   \thesaurus{03
	      (03.13.4;
	       11.06.1;
	       11.11.1;
	       12.03.4;
	       02.08.1)}
   \title{Placing stars within cosmological simulations}

   \author{F.R.Pearce\thanks{e-mail: F.R.Pearce@durham.ac.uk}}

   \institute{Physics Department, University of Durham, Durham, UK \\
\and  Astronomy Centre, University of Sussex, Falmer,Brighton, UK	
             }

   \date{Received -; accepted -}

\maketitle
\begin{abstract}
 \input abstract
\end{abstract}
\input intro

\input sim

\input result

\input conc

\input ack

\input ref
\end{document}

%% file: refer.tex
\def\P3M{\hbox{$P^{3}M$}}

\def\AP3M{\hbox{$AdP^{3}M$}}

\def\cc2{c2}
\def\cc3{c3}
\def\cc4{c4}
\def\cc{c}

\def\circ{circular}

\def\s{r_s}

\def\ApJ{ApJ }

\def\ApJS{ApJS }

\def\ARAA{ARA\&A }

\def\MN{MNRAS }

\def\PhD{PhD thesis }

\def\prep{in preparation }
\def\press{in press }

\def\ref{\par \noindent \hang}
\def\etal{{\it et al.\thinspace}}
\def\eg{{\it e.g.\ }}

\def\spose#1{\hbox to 0pt{#1\hss}}
\def\approxlt{\mathrel{\spose{\lower 3pt\hbox{$\sim$}}
	\raise 2.0pt\hbox{$<$}}}
\def\approxgt{\mathrel{\spose{\lower 3pt\hbox{$\sim$}}
	\raise 2.0pt\hbox{$>$}}}

\def\<{\thinspace}
\def\s{\hbox{\phantom{5}}}	

\def\boxit#1{\vbox{\hrule\hbox{\vrule\kern3pt\vbox{\kern3pt
          #1 \kern3pt}\kern3pt\vrule}\hrule}}

%% file: units.tex
\def\km{{\rm\thinspace km}}
\def\kpc{{\rm\thinspace kpc}}

\def\Mpc{{\rm\thinspace Mpc}}
\def\Msun{\hbox{$\rm\thinspace M_{\odot}$}}

\def\s{{\rm\thinspace s}}

\def\h50{\hbox{$\rm\thinspace h_{50}$}}
\def\h50m1{\hbox{$\rm\thinspace h_{50}^{-1}$}}

\def\kmpspMpc{\hbox{$\km\s^{-1}\Mpc^{-1}\,$}}

%% file: abstract.tex
I investigate the process of converting gas into stars within the
framework of a standard cosmological model. By examining the set of
objects grown in a combined N-body plus smoothed particle
hydrodynamics simulation with those obtained in similar models where
some of the cold, dense gas was replaced by collisionless ``star''
particles I show that it is possible to make this substitution
without affecting the subsequent gas cooling rate.  
With even the most basic star forming criteria the
masses of isolated objects are nearly identical to the mass of
cold, dense gas found within the same objects in a non-star forming run.

No evidence is found to support the contention that converting gas
into stars might affect the amount of cold gas obtained in a
simulation by retarding the cooling rate within those objects where
stars have already formed.  
In practice, because cold gas can be reheated by shocks but stars
remain as such whatever happens
the masses of the largest objects found in the
star forming runs are generally higher than those in the standard run.

Finally, I demonstrate that an excellent match to the observed star formation
rate can be achieved with even a very basic star formation
prescription.

%% file: intro.tex
\section{Introduction}

Collisionless cosmological simulations have for several years been
expanding our knowledge of how the large scale structure in
the Universe formed. The difficulty lies in comparing the dark matter
of the models with the observable Universe, where we see starlight from
galaxies and X-rays from hot gas.  Gas is a dissipational medium;
tightly bound structures form which survive the violent evolution of
larger objects. When galaxies fall together to form clusters their
dark matter halos become stripped and merge together, leaving behind
several distinct, bright galaxies within a single dark matter halo.
Collisionless simulations lack the dissipational processes that
occurred as visible galaxies cooled and formed stars which makes the
process of comparing the results to the observations difficult.

Recently several codes have been written that are capable of including
a dissipating gaseous component (Katz, Hernquist \& Weinberg 1992,
Navarro \& White 1993, Evrard, Summers \& Davis 1994, Tsai, Katz \&
Bertschinger 1994, Couchman, Thomas \& Pearce 1995, Cen \& Ostriker
1996, Frenk \etal 1996, Steinmetz 1996). Most of these groups employ
smoothed particle hydrodynamics (SPH) to follow the gas (Monaghan
1992). This allows non-adiabatic heating through shocks and cooling via
radiation to be included in the cosmological models and a direct
comparison can be attempted between the regions of cold, dense gas and
the observed distribution of galaxies.

Here I examine the formation of objects within a cosmological
volume. Typically a Milky Way like galaxy will contain 100 or so
gas particles with the mass resolution I employ. With this number
of particles it is not possible to recover the internal
dynamics and structure of the objects but
reasonable distribution and multiplicity functions can be
obtained. 

For a variety of reasons, not least because it saves computational
effort, it is useful to convert the gas within cold, dense
clumps (or ``galaxies'') into ``star'' particles, which from
that point on behave like collisionless dark matter particles.
This can be achieved in a variety of ways, with varying degrees of
physical motivation. Here I will attempt to show that perhaps the
simplest possible choice of assumptions leads to a stable result and
so more complicated schemes are perhaps unnecessary. Throughout this
paper the term ``star'' refers to a particle that was once subject to
gas forces but has since been converted into a collisionless particle
of the same mass. I do not consider any additional physical processes
which might affect a stellar population such as supernova feedback or
metal enrichment as these will be discussed in detail elsewhere.

Within this paper I show that one successful way of incorporating
star formation within a cosmological simulation without affecting the
masses of the final objects is to use a combined density and
temperature cut. This alleviates the worry that the action of
converting a gaseous particle into a collisionless one might retard
subsequent cooling by lowering the local gas density.  In practice
this is not the case and apart from dynamical considerations we are
free to equate the mass of stars found with the mass of cold, dense
gas that would have been obtained in a non-starforming
simulation.

I then compare the star formation rate obtained with my preferred
method against that observed by Madau (1996). The good agreement 
between the observations and the simulations demonstrates that
reproducing the observed star formation rate is in practice
straightforward and is a natural consequence of hierarchical clustering.

%% file: sim.tex
\section{Simulations}

I have run my simulations using a parallel adaptive
particle-particle, particle-mesh plus smoothed particle hydrodynamics
code (Pearce \& Couchman 1997)
implemented in CRAFT, a directive based parallel Fortran developed by
Cray.  It can be run in parallel on a Cray T3D or serially on a single
processor workstation and is essentially identical in operation to the
publicly released version of Hydra (Couchman, Pearce \& Thomas 1996)
which is described in detail by Couchman, Thomas \& Pearce (1995).
This work has been carried out as part of the programme of the 
Virgo Consortium, a group of astrophysicists interested
in large cosmological simulations. 

The three simulations presented here 
were of an $\Omega_0=1$, standard cold dark matter
Universe with a boxsize of $25 h^{-1}\Mpc$. I take $h=0.5$ throughout
this work, equivalent to a Hubble constant of $50 \kmpspMpc$.  
The baryon fraction, $\Omega_b$ was set from
nucleosynthesis constraints, $\Omega_bh^2 = 0.015$ (Copi, Schramm \&
Turner 1995) and I assume a constant gas metallicity of 0.5 times the
solar value. Identical initial conditions were used in all cases,
allowing a direct comparison to be made between the objects
formed. Useful parameters for the runs are listed in table~1.

The
initial fluctuation amplitude was set by requiring that the model
produced the same number density of rich clusters as observed
today. To achieve this I take $\sigma_8 = 0.6$, the present day
linear rms fluctuation on a scale of $8h^{-1}\Mpc$ (Eke, Cole \& Frenk
1996, Viana
\& Liddle 1996).  Each model began with $64^3$ dark matter particles
each of mass $3.1 \times 10^{10} \Msun$ and $64^3$ gas particles of
mass $1.9
\times 10^9 \Msun$, smaller than the critical mass derived
by Steinmetz \& White (1996) required to prevent 2-body heating of the
gaseous component by the heavier dark matter particles.  I employ a
comoving Plummer softening of $25 h^{-1}\kpc$, which is typical
for cosmological simulations but much larger than required
to accurately simulate the dynamics of galaxies in dense environments.
For a Plummer softening of $25 h^{-1}\kpc$ the effective softening
is around $50 h^{-1}\kpc$ whilst elliptical galaxies have scalelengths
of around $4 h^{-1}\kpc$. This dramatic increase in the cross section
of the objects will artificially enhance the merger rate and make
tidal stripping much more effective.

With only $2 \times 64^3$ particles converting the dense gas into
collisionless particles does not result in a large time saving.
The total number of steps taken is not reduced by much because
the dissipation that occurred as the gas cooled to
form dense knots produces tightly bound clumps of stars which require
small timesteps if they are not to be disrupted. A modest saving is
seen in the CPU time required per step as for stars expensive SPH calculations
no longer need to be carried out.
The full benefit of converting
gas into stars only really becomes apparent within
larger $2 \times 128^3$ particle simulations where 
shorter timesteps are observed (Pearce \etal 1997) if
star formation is not employed because the volume contains larger
objects which form earlier.

Before star formation can be introduced into a model an adequate
treatment of gas cooling is required. In this paper I use the
analytic form of Thomas \& Couchman (1992).  I have implemented
the tabulated cooling functions of Sutherland \& Dopita (1993)
and no significant differences to the results presented here
are obtained. In practice the results
presented in this paper are not sensitive to the precise form of the
cooling function and should hold for all such functions.

The gas ends up in three phases; there is a hot phase, where gas sits in 
the potential wells formed by the larger objects, and two cold phases,
one formed by the collapsed objects and the other formed by gas that
has yet to collapse which occupies the void regions. Table~1 lists
the percentage of the gas which resides in each of these phases at a redshift
of zero.
These values are in agreement with those obtained by Evrard, Summers
\& Davis (1994) at a redshift of 1 (the end of their simulation),
where I obtain values of 24 percent hot gas and 72 percent cold gas in voids
in all three simulations, the remaining 4 percent is in the form of cold,
collapsed gas or stars.

\begin{table}
 \input table1
 \caption{Parameters for the 3 runs. These are the run
 without any star formation, one employing a density plus a temperature cut, 
and finally a density, temperature and small probability cut. The
columns list the number of star
particles at the end (or for the no star case the number of
cold, dense particles which would have formed stars if a density
plus a temperature cut were employed on the final step), 
the number of groups found by the group finder at the end of the simulation, 
the percentage of the stars that are within objects defined by the group
finder, the percentage of the gas that is above $10^5K$, the percentage
of gas that is below $10^5K$ and at low density, the number of system steps
required and the total CPU time for the entire run in hours on  
64 T3D processors.}

 \label{timediag}
\end{table}

\subsection{Star formation details}

Stars are formed within the models in two slightly different ways.  In
both cases once a gas particle has satisfied the relevant criteria its
type is simply changed from ``gas'' to ``star'' and from then on the
particle behaves like a collisionless dark matter particle of the same
mass, only experiencing the force of gravity.  This ``instantaneous''
conversion of a gas particle into a star is deliberately crude,
reducing the number of free parameters to a minimum and is similar to
that employed by Summers (1993). I do not consider schemes which
involve spawning additional particles (\eg Katz 1992, Navarro \& White
1992), redistributing mass or dual identity ``schizophrenic''
particles (Mihos \& Hernquist 1994).  All these schemes may increase
the tunability of the star forming process but as I will later show
they are not required at least in a cosmological situation where
the poor mass resolution makes any model of star formation very crude.

In the first model a star is formed if a gas particle has a density
exceeding $550$ times the mean density in the box and has a
temperature below $10^5K$, converting 11 percent of the gas into stars
by the end of the simulation ($\Omega_\star=\Omega_b \times 0.11=0.006$).  This
is close to the observed stellar mass fraction, $\Omega_\star =
0.004$ (Peebles 1993) despite the poor mass resolution of these
simulations. This is another demonstration of the well known cooling
catastrophe (White \& Frenk 1991).  
The second method is the same as the first except that a
gas particle only has a $5\%$ chance of becoming a star on any one
step.  An overdensity of $550$ is quite low compared to the threshold
used by other authors (\eg Navarro \& White 1992) but I am forced to
adopt this value because of my poor mass resolution and large
softening which causes the maximum reliable overdensity obtained within the
simulation to be around 2000.  In practice cold gas in this regime
has little pressure support and so rapidly increases in density.

Within a cosmological simulation it seems non-sensical to adopt
a physical density criterion as the basis for a star formation
algorithm. The star forming regions are characterised 
as collapsed
clumps of cold, dense particles where the important factor is not
the physical density but the overdensity. At a high enough
redshift the entire Universe will be at a density above whatever
physical threshold is employed, forcing the implementation of
an overdensity constraint to prevent star formation
at early times. If the star forming
region itself could be resolved within a simulation then a physical
density threshold might be useful, but when an entire galaxy is
only barely resolved such a threshold seems at best problematic. 

\subsection{Identifying groups}

Within a simulation volume it is useful to be able to reliably
identify a catalogue of all the bound objects. In practice this is a
tricky procedure because some degree of merging and disruption of the
object set will be taking place at any given time and a binary merger
can dramatically affect the position of any given object within the
catalogue. The presence of a cooling gaseous component reduces the
problem of identifying groups because both density and temperature
information is available.  The objects being sought are characterised as
clumps of cold, dense particles.  In principle both density and
temperature information could be calculated for the dark matter
component but the lack of dissipation and large scatter reduces the
usefulness of the local velocity dispersion although others have
used the local dark matter density as an aid to defining groups (Gelb
\& Bertschinger 1994).

To define groups I use the procedure of Thomas \& Couchman
(1992).  By extracting all the gaseous particles which are
simultaneously both in a dense region and have a temperature below
$10^5K$ a reduced set of positions is obtained which can then be passed to a
friends-of-friends algorithm. When the maximum linking length is small
the particles within each clump are linked together. At higher values
the recovered object set remains static until neighbouring clumps
begin to be linked. This allows us to use a maximum linking length
of $60h^{-1}\kpc$, a value larger than normal because of the highly
reduced set of points that are passed to the finder.  In practice
the maximum linking length can vary over a large range and essentially
the same object set be recovered.  

Once star formation is implemented the stars
themselves provide an additional method for making the initial
selection of those particles which might be in groups. 
For the runs which contain star particles the friends-of-friends
finder uses the star positions to produce the group catalogue
with the same linking parameters as for the gas runs. 
With stars the isolated objects are still easy to recognise
and define but within dense regions things are much more messy.
Clumps of stars are often non-spherical due to tidal distortions
and the whole cluster is permeated by a diffuse background light
that originates from small systems that have been completely
disrupted. The central part of this halo is linked with the
mass of the central object but in practice few galaxies
are linked together because the stellar systems remain 
centrally concentrated. In total 86 percent of the stars end up in the
object set produced by the group finder. For the gas only run a
much higher value of 97 percent is achieved, simply demonstrating
the messier nature of runs which include stars.
Practically, the merger and halo effects could be alleviated by reducing the
value of the linking length used to define the stellar clusters
but this would further reduce the percentage of stars lying
within resolved objects.

%% file: table1.tex
\begin{tabular}{lccccccc}
Run & $N_\star$ & $N_{g}$ & $\%_{g}$ & $\%_{h}$ & $\%_{c}$ & Steps & cpu  \\
No stars & 31369 & 315 & 97 & 46 & 43 & 1897 & 20.8 \\
$\rho + T$ & 29740 & 327 & 86 & 45 & 44 & 1732 & 16.6 \\
$\rho + T + \%$ & 29033 & 320 & 87 & 45 & 44 & 1822 & 18.1 \\
\end{tabular}

%% file: result.tex
\section{Results}

\subsection{Multiplicity function}

\begin{figure*}
 \centering
\psfig{file=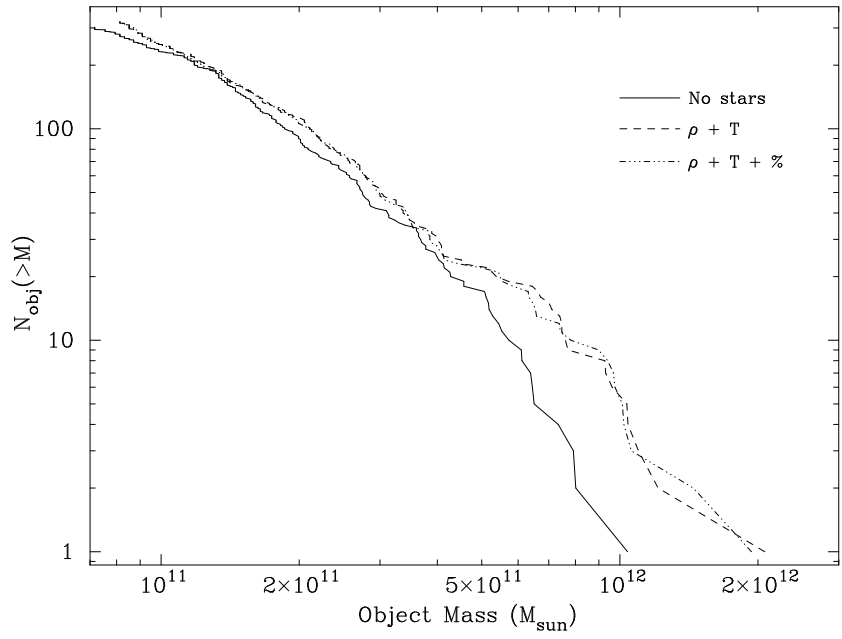,height=9cm}
 \caption{The multiplicity function}
 \label{massob}
\end{figure*}

The group finder described above produces a catalogue of objects
ordered by mass. The effect on the multiplicity function of forming
stars using my different prescriptions is shown in figure~1. 

The gas particles being considered for conversion into a star 
should always be both dense and
cold, residing within collapsed objects. 
If only a density threshold were employed it would be necessary to choose
a value low enough to allow small, isolated clumps to form stars but also high
enough to prevent hot gas particles in the cluster halo being
converted.  The extra computational
effort of examining the temperature is small and removes the problem
of dense, hot cluster halos being converted into stars. 
As an artifact of the SPH approach it is
also possible for a hot halo particle to overestimate its local
density if it passes close to the central cold gas clump. With just a
density cut these particles may be erroneously flagged as stars.

The most massive stellar objects formed within the starforming runs 
are generally twice as massive
as those in the standard run. 
The clustered regions where these large objects reside are not being
overmerged by the group finder as near the
largest object the group finder recovers 13 objects in the standard
run as opposed to 12 in the stellar run. One tiny object near the
centre of the star cluster has been merged, raising the central object mass
by less than one percent. This shows that the central object really
does contain twice as many stars as the amount of cold gas in the
non-starforming run. 
Dense, cold gas provides little pressure support and so its removal
should not affect the cooling rate of the halo, as demonstrated by the
observation that the central temperature of the hot halo
when stars are present is only 1.2 percent below that of the standard 
(no star) run and the percentage of gas in the hot phase is almost
constant between the three runs.

The reason for the more massive stellar objects is that small objects
can be tidally disrupted as they fall into a cluster
(Summers 1996, Navarro \& Steinmetz 1997).
When the stellar clumps become tidally disrupted the stars 
remain stars and form a diffuse halo, some of which gets
attributed to the central object. 15 percent
of those particles which are both cold and dense enough to have been
called stars at a redshift of 0.5 do not remain cold and dense at the
end of the simulation. These particles predominantly lie in objects
that are in the process of infalling into the larger clusters within
the simulation volume and so the disruption process is heavily
biased. This disruption should be seen as a 
numerical artifact due to the large softening length employed. With a
smaller softening length both the cold, dense clumps and the stellar 
associations would have been
allowed to dissipate further as they formed and perhaps survive passage
through the harsh cluster environment.

\subsection{Star formation history}

\begin{figure*}
 \centering
 \psfig{file=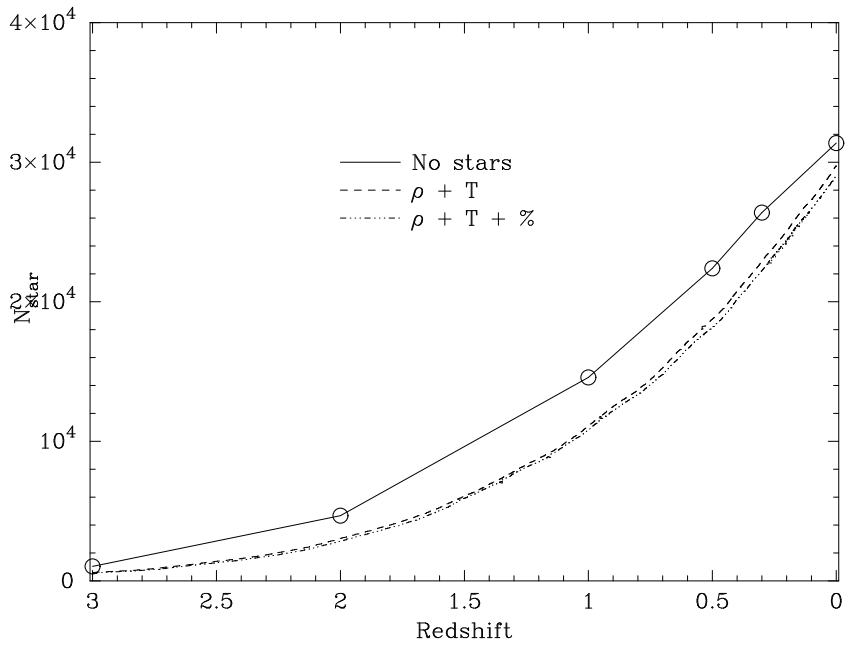,height=9cm}
 \caption{The star forming history}
 \label{starhist}
\end{figure*}

\begin{figure*}
 \centering
 \psfig{file=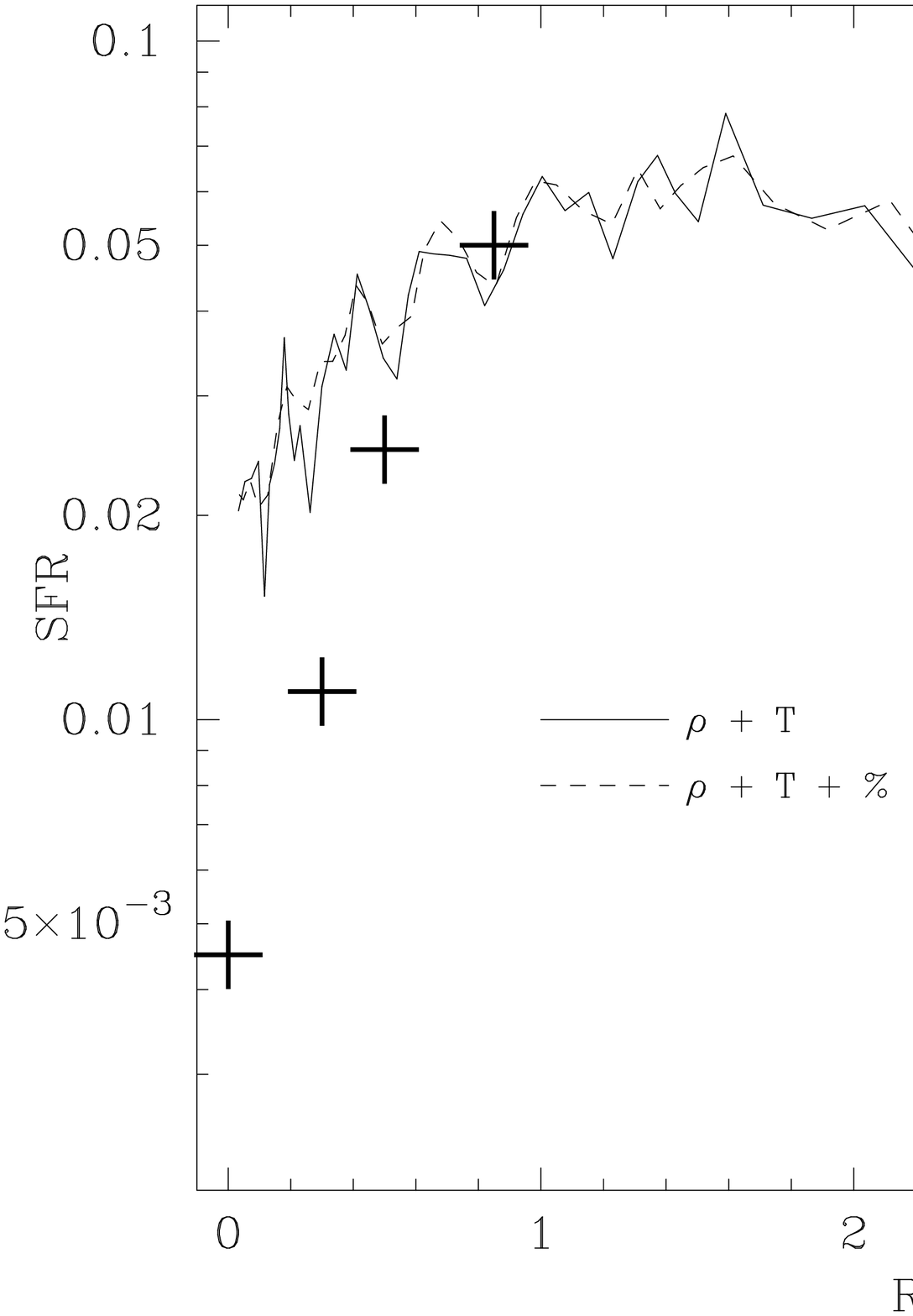,height=9cm}
 \caption{The star formation rate in $M_\odot/{\rm yr}/{\rm Mpc}^3$ versus redshift
for the simulations with a density and temperature cut with and
without a delay via a percentage chance. The large crosses show the
observed points taken from Madau (1996)}
 \label{starrate}
\end{figure*}

The redshift at which stars were created is shown in figure~2.  The
first ``stars'', formed in the precursor objects of what become
the central galaxies of the largest clusters, 
appear around a redshift of 4. This initial
formation epoch should be treated with extreme caution because if the
resolution were to be increased then these first stars would be seen
even earlier. This is the well known ``cooling catastrophe'' where
more and more gas cools as the resolution is progressively
increased (White \& Frenk 1991). 
The formation time of the earliest objects also
depends upon the size of the volume being simulated; bigger boxes have
more room to fit in a high sigma peak of the initial fluctuation
spectrum, providing a site where a large cluster will eventually form. If
the resolution is fixed bigger boxes will form their first objects
earlier.

As can be seen the standard, non-star forming run always has more
particles which would have qualified as stars if the same density and
temperature cut were employed as for the star forming runs.
This is despite the disruption of objects mentioned in
the previous section.  The reason for this discrepancy is that 
the star forming algorithm is discrete, but the SPH variables
used to measure the state of the gas are averages over some
number, $N_{sph}$, of particles (here $N_{sph} = 32$).
For a single star to be formed there must be $N_{sph}$ particles
close to the star formation threshold and so a zero
point offset has to be introduced if the object masses are to be
compared.  To compensate for this effect I add $\alpha N_{sph}$ onto the mass
of all the stellar groups, where $\alpha$ is a parameter between 0 and
1. For simplicity I take $\alpha=1$ in what follows. 

In figure~3 I plot the star formation rate in $M_\odot/yr/Mpc^3$ 
versus redshift with the observed results from
Madau (1996) overlayed. Clearly stars are being overproduced at late times
(after a redshift of 1) but this may just be a consequence of the
cosmological model employed (standard CDM has lots of evolution
at late times). This demonstrates that any star formation
formula that is ultimately based upon the gas density and is
normalised to produce the correct total mass of stars should
be expected to fit the observed star formation rate very well.

As can be seen from figures~1, 2 \& 3 the runs with a percentage chance of
forming a star on any one step produce almost identical results to 
their counterparts. 
The initial motivation for introducing a percentage chance to delay
star formation was to
assist gas cooling by providing a ``seed'' of cold gas which was
not immediately converted into stars onto which
more gas could accumulate. The success of the model with a combined
density and temperature cut demonstrates that this concern was
unfounded and that delaying star formation via a random chance
is an unnecessary complication that in practice simply raises
the density threshold at which stars are formed because during
the delay introduced the clump collapses further. This has the
effect of producing more tightly bound clumps which may then
survive subsequent tidal disruption but in practice has little 
effect upon the simulation. This lack of effect occurs for the same reason
that a zero point offset is required; at any one time $N_{sph}$
particles are close to the star formation threshold so the addition of
a few extra particles above the threshold has little effect.

\subsection{Direct comparison}

In figure~4 I show a comparison between the masses of the objects
found by the group finder in the standard non-star forming run and
the masses of the same objects found in my preferred star
forming run, where both density and temperature cuts were employed
but without a percentage chance (which has little effect).

\begin{figure*}
 \centering
\psfig{file=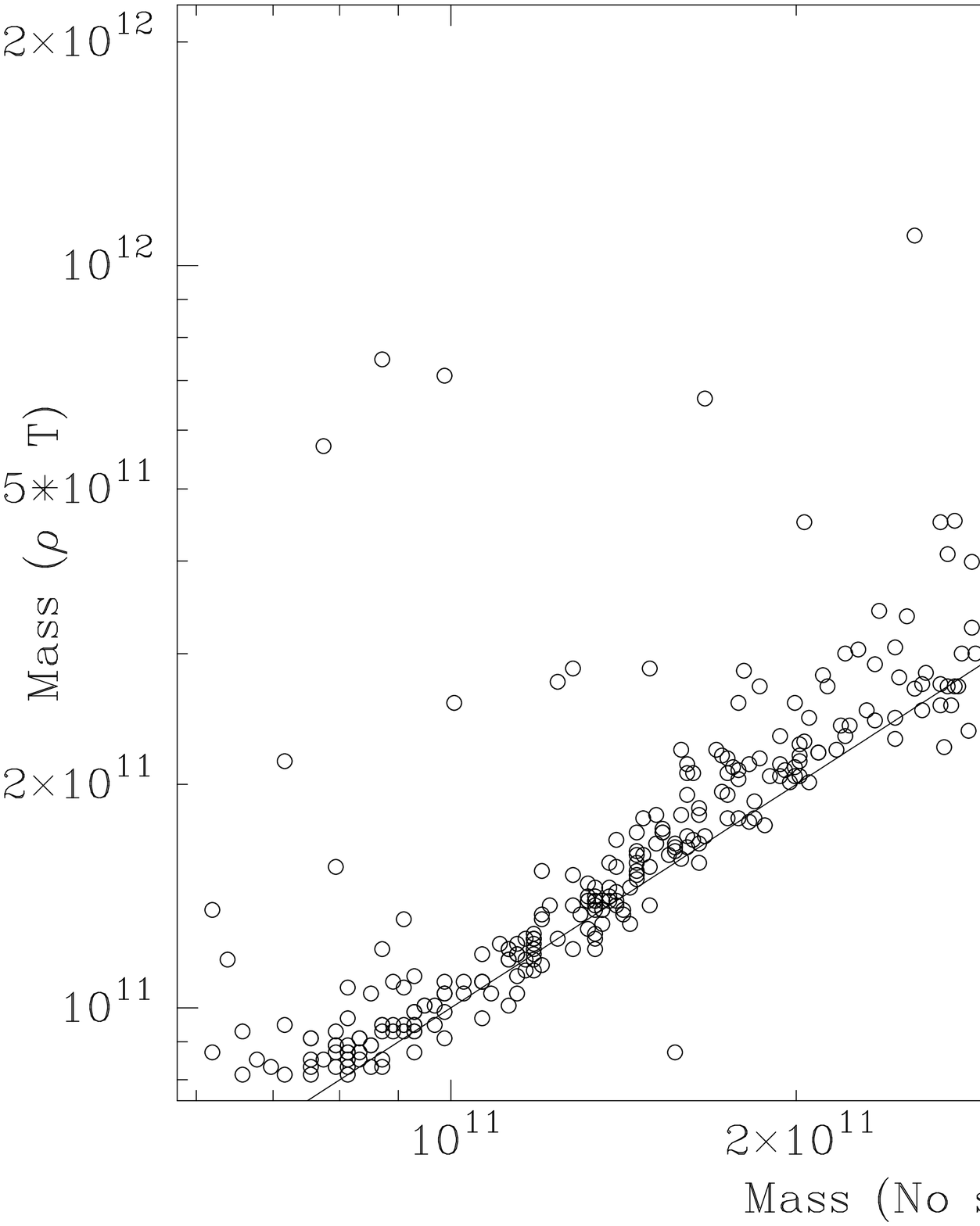,height=14cm}
 \caption{Comparison of the masses of objects with and without star
formation. The line shows where identical masses would fall.}
 \label{masscom}
\end{figure*}

The correspondence in mass between the objects in the two runs 
is very good with only a small scatter. At the high mass end
the masses of the objects in the star forming run are in fact higher
than the equivalent object in the run without stars. This is because
some objects falling into the cluster halos are disrupted and
dispersed. When stars are present these subsequently add to the mass
of the central object but without stars the gas is reheated and
forms part of the hot cluster halo.
With a smaller gravitational softening these objects would be
more tightly bound and harder to disrupt and so the object
masses in the non-star forming run would have been higher.

The few outliers in the upper left quadrant of the figure are
due to objects within the star forming simulation that
have been linked together by the group finder. This is because 
with stars the objects often have a small ``halo'' because
they cannot dissipate further once stars have been formed and
subsequent mergers and tidal torques lead to messier shapes,
forming bridges and spurs which sometimes cause the group finder
to link together separate objects. The paucity of objects in this region 
of the plot clearly demonstrates that overmerging 
is not a problem with these simulations.

%% file: conc.tex
\section{Conclusions}

The main conclusion of this work is that replacing cold, dense
gas with stars within a cosmological simulation
can be done with relative ease and to some degree of accuracy without 
affecting the subsequent cooling rate of the gas. In principle
all that is required is that the gas is dense enough but in
practice a temperature threshold should also be employed to
ensure that hot halo particles are not converted to stars erroneously.

Introducing a percentage chance to delay star formation raises the
density threshold and has no affect on the masses of the objects
obtained.  This and more complicated schemes such as the gradual
conversion of a gas particle into a star might be useful if additional
physical processes such as feedback are considered or if a smoother
star formation rate is desired but are not
required in principle.

It is a fundamental property of the SPH method that if one particle
has a certain density and temperature $N_{sph}$ particles should
have similar properties. This leads to a zero point offset in
the masses of the stellar objects when compared to the mass of
cold, dense gas in a non-star forming run.

Matching the star formation rate obtained within a simulation to
that observed is a straightforward exercise with good fits being
obtained with even the most basic star formation formulae. This
is because the general trend of the observed values, a small rate
at early times rising slowly to a peak around a redshift of 1
followed by a decline just mimics the effects of structure
formation in any underlying cosmology. 

Although the introduction of star formation allows simulations to be
carried out more rapidly the simulator should be aware of several
additional processes it introduces. Isolated objects are well
reproduced whether or not star formation is employed but within
dense regions additional dynamical processes are taking place.
Obviously the dynamics of a gaseous clump will be different
from a collisionless object in an environment where ram pressure stripping
and merging are important. Small cold gas clumps can be disrupted
as the cluster forms, dissolving into the general cluster halo. The 
analogous stellar systems will also be tidally disrupted but the
stars themselves will survive and contribute to a diffuse
cloud of stellar particles that pervades the larger groups. Gas clumps
merge quite easily whereas star clumps tend to produce messy
associations because once a star has been formed no subsequent
dissipation can take place and so reheating events can prove to be a
problem. 

In summary, converting gas into stars within a cosmological N-body
simulation is a viable option and even a basic prescription
works well for isolated objects. However, in regions where
tidal torques and merging are significant the differences
between collisionless systems and collisional, dissipating
gaseous systems should be carefully considered and if possible
a parallel non-star forming simulation should be carried out.

%% file: ack.tex
\subsection*{Acknowledgments}

FRP was an EPSRC PDRA working for the Virgo Consortium.  The work
presented in this paper was carried out as part of the programme of
the Virgo Supercomputing Consortium using computers based at the
Computing Centre of the Max-Planck Society in Garching and at the
Edinburgh Parallel Computing Centre.  This paper benefited greatly
from discussions with numerous people, not least Peter Thomas, Adrian
Jenkins, Carlos Frenk, David Weinberg, Scott Kay, John Baker, Hugh
Couchman, Julio Navarro, Gus Evrard and Simon White.

%% file: ref.tex
\subsection*{References}

\noindent Cen, R., Ostriker, J., 1996, \ApJ 464, 270

\noindent Copi, C. J., Schramm, D. N., Turner, M. S., 1995, \ApJ 455,
95


\noindent Couchman, H. M. P., Thomas, P. A., Pearce, F. R., 1995,
\ApJ 452, 797

\noindent Couchman, H. M. P., Pearce, F. R., Thomas, P. A., 1996,
astro-ph/9603116





\noindent Eke, V. R., Cole, S., Frenk, C. S., 1996, \MN  282, 263

\noindent Evrard, A. E., Summers, F. J., Davis, M., 1994, \ApJ  422, 11

\noindent Frenk, C. S., Evrard, A. E., White, S. D. M., Summers,
F. J., 1996, \ApJ  472, 460


\noindent Gelb, J. M., Bertschinger, E., 1994, \ApJ  436, 467 



\noindent Katz, N., 1992, \ApJ  391, 502

\noindent Katz, N., Hernquist, L., Weinberg, D. H., 1992, \ApJ  399, 109

\noindent Mihos, C. J., Hernquist, L., 1994, \ApJ  437, 611

\noindent Madau, P., 1996, astro-ph/9612157

\noindent Monaghan, J. J., 1992, \ARAA 30, 543

\noindent Navarro, J. F., White, S. D. M., 1993, \MN  265, 271

\noindent Navarro, J. F., Steinmetz, M., 1997, \ApJ  478, 13

\noindent Pearce, F. R., Couchman, H. M. P., 1997, NewA 2, 411

\noindent Peebles, P. J. E., 1993, Principles of Physical Cosmology, Princeton
University Press




\noindent Pearce, F. R., Thomas, P. A., Jenkins, A. R. \etal, 1997, \prep

\noindent Steinmetz, M., 1996, \MN  278, 1005

\noindent Steinmetz, M., White, S. D. M., 1996, \press

\noindent Summers, F. J., 1993, \PhD, University of California

\noindent Summers, F. J., 1996, astro-ph/9602119

\noindent Sutherland, R. S., Dopita, M. A., 1993, \ApJS 88, 253


\noindent Thomas, P. A., Couchman, H. M. P., 1992, \MN  257, 11

\noindent Tsai, J. C., Katz, N., Bertschinger, E., 1994, \ApJ  423, 553

\noindent Viana, P. T. P., Liddle, A. R., 1996, \MN  281, 323

\noindent White, S. D. M., Frenk, C. S., 1991, \ApJ  379, 52